\documentclass[3p,times,procedia]{elsarticle}
\usepackage{nupha_ecrc}

\volume{00}

\firstpage{1}

\journalname{Nuclear Physics A}

\runauth{}

\jid{nupha}

\jnltitlelogo{Nuclear Physics A}

\usepackage{amssymb}
\usepackage[figuresright]{rotating}

\begin{document}

\begin{frontmatter}



%% Instructions from Editor: Please use the following \dochead only in the preprint version (e-print arXiv etc.); 
%% use empty \dochead{} when submitting to Nuclear Physics A!
\dochead{XXVIIth International Conference on Ultrarelativistic Nucleus-Nucleus Collisions\\ (Quark Matter 2018)}
%% e.g. \dochead{17th International Conference on Dynamical Processes in Excited States of Solids}

%\title{Electromagnetic probes of the early stage of heavy ion collisions}
\title{Photon radiation from heavy-ion collisions in the $\sqrt{s_{NN}}=19-200$~GeV regime}


\author[label1]{Charles Gale}
\author[label1]{Sangyong Jeon}
\author[label1]{Scott McDonald}
\address[label1]{Department of Physics, McGill University, 3600 University Street, Montr\'eal, QC, H3A 2T8, Canada}
\author[label2]{Jean-Fran\c cois Paquet}
\address[label2]{Department of Physics, Duke University, Durham, NC 27708, USA}
\author[label3]{Chun Shen}
\address[label3]{Department of Physics, Brookhaven National Laboratory, Upton, New York 11973-5000}

%\address{}

\begin{abstract}
We present calculations of prompt and thermal photon production in Au-Au collisions at $\sqrt{s_{NN}}=19-200$~GeV. We discuss features of the spacetime profile of the plasma relevant for electromagnetic emission. We highlight how the suppression of prompt photon production at low $\sqrt{s_{NN}}$ can provide a window to measure thermal photons in low collision energies.
\end{abstract}

\begin{keyword}
Electromagnetic radiation \sep heavy-ion collisions \sep quark-gluon plasma \sep RHIC Beam Energy Scan

\end{keyword}

\end{frontmatter}

% \linenumbers

\section{Introduction}
\label{sec:intro}

The strongly-coupled quark-gluon plasma produced in heavy ion collisions extends approximately 5 to 10 fm in the plane transverse to the beam axis. This volume of the plasma is sufficient to produce a measurable amount of thermal electromagnetic (``black-body'') radiation. On the other hand, the plasma's dimensions are sufficiently small for this electromagnetic radiation to escape the plasma with negligible rescattering (see Ref.~\cite{Paquet:2016pnt} and references therein). This makes photons invaluable probes of entire spacetime dynamics of heavy ion collisions.
%This implies that photons and dileptons can be used to probe every stages of heavy ion collisions, including the early dynamics of the plasma that is more difficult to probe with typical hadronic observables.

%In this work, we use photons to study the early stage of heavy ion collisions  in the 

In this work we study photon production in Au-Au collisions in the $\sqrt{s_{NN}}=19-200$~GeV collision energy regime. This corresponds to most of the collision energies probed by the RHIC Beam Energy Scan (Phase 1). 

%At high $\sqrt{s_{NN}}$ ($\sqrt{s_{NN}}=200$~GeV and higher), the initial energy deposition in heavy ion collisions can be treated as nearly instantaneous, at least in the midrapidity region~\cite{Shen:2017bsr}. This is not the case for lower $\sqrt{s_{NN}}$ collisions, which thus require a more complex model of energy deposition over an extended period of time~\cite{Shen:2017bsr}. In this work we show the power of photons in differentiating energy deposition scenarios.

\section{Evolution of spacetime profile with $\sqrt{s_{NN}}$}
\label{sec:evol}

The spacetime evolution of the strongly-coupled quark-gluon plasma is described with $3+1$D relativistic viscous hydrodynamics. The conservation equation for the net baryon density is solved alongside energy-momentum conservation and the Israel-Stewart-type relaxation equation for the shear stress tensor~\cite{Denicol:2018wdp}. Energy deposition is performed by adding sources to the conservation equations, as described in Ref.~\cite{Shen:2017bsr}. This dynamical initialization for the hydrodynamics allows for energy and baryon number to be deposited over an extended period of time, which is the expected scenario in lower energy collisions. A more complete description of the hydrodynamic model can be found in Refs.~\cite{Shen:2018pty,Shen:2017fnn}.

At higher collision energies, the baryon current conservation has a relatively small effect on the plasma evolution and can be neglected. In this case the dominant factor determining photon production is the temperature profile of the quark-gluon plasma: how much spacetime volume of plasma is radiating, and what is the temperature distribution of this four-volume. The main difference with lower $\sqrt{s_{NN}}$ collisions is the increasing importance of the baryon chemical potential $\mu_B$, which makes it necessary to account for both the temperature and baryon chemical potential profile.

%The momentum distribution of these produced photons is moreover affected by the flow velocity distribution of the plasma. As for the viscosity, it has a dual effect. First it enters into the fluid dynamic equation describing the evolution of the plasma and affect its space-time evolution. Second it modifies the local rate of photon emission.

\begin{figure}
	\centering
	\includegraphics[width=0.44\linewidth]{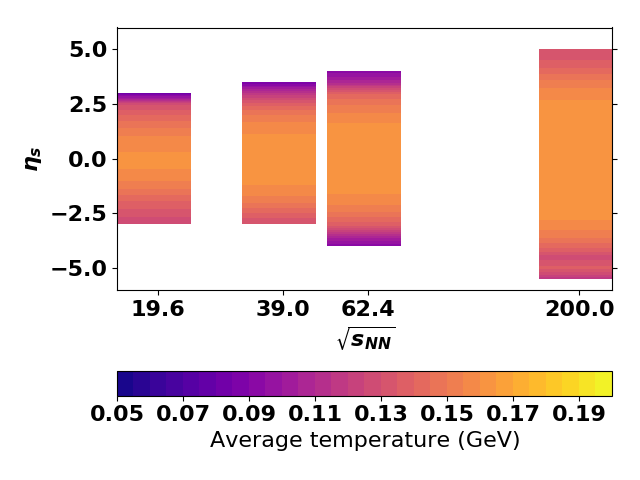}
	\includegraphics[width=0.44\linewidth]{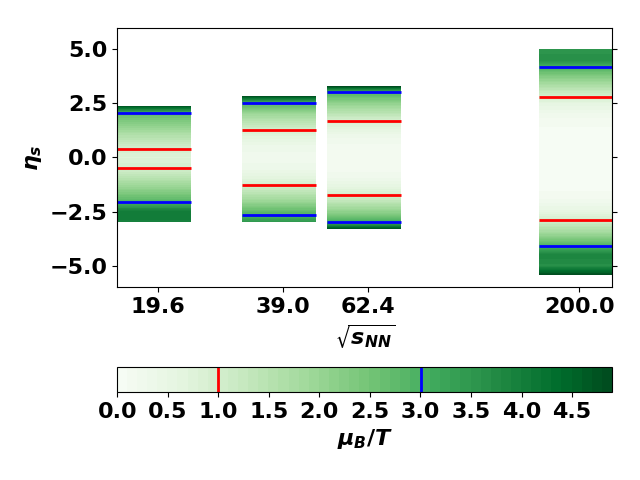}
	\caption{Average temperature (left) and average $\mu_B/T$ (right) as a function of center-of-mass energy $\sqrt{s_{NN}}$ and spatial rapidity $\eta_s$ for $0-5$\% Au-Au collisions. Markers were added at $\mu_B/T=1$ (red) and $\mu_B/T=3$ (blue) on the right-hand plot for reference.}
	\label{fig:rapidity_cm}
\end{figure}

The evolution of the temperature profile at high collision energies has been investigated in multiple publications (see e.g. Refs.~\cite{Chatterjee:2013naa,Shen:2013vja,Paquet:2015lta}), predominantly in boost-invariant ($2+1$D) calculations that are appropriate for midrapidity observables. In this section we show the dependence of the temperature and baryon chemical potential profile on spatial rapidity of the plasma and on the center-of-mass energy of the collisions. Averages of the temperature and of $\mu_B/T$ are calculated and shown as a function of spatial rapidity $\eta_s$ and center-of-mass energy  $\sqrt{s_{NN}}$ in Figure~\ref{fig:rapidity_cm}.

The averages are computed with the definition 
$
\langle \ldots \rangle=\int_{\epsilon(X)>0.16\textrm{ \footnotesize GeV/fm}^3} d^4 X \ldots / \int_{\epsilon(X)>0.16\textrm{ \footnotesize GeV/fm}^3} d^4 X
\label{eq:average}
$
with ``$\ldots$'' being either $T(X)$ or $\mu_B(X)/T(X)$. The energy density cut-off $\epsilon(X)>0.16\textrm{ GeV/fm}^3$ corresponds to a temperature of T=145~MeV at $\mu_B=0$. A single event with 0-5\% centrality is used for each $\sqrt{s_{NN}}$. Note that the averages are not perfectly symmetrical in spatial rapidity because of physical event-by-event fluctuations in the energy deposition.

Figure~\ref{fig:rapidity_cm} summarizes a large amount of information about the space-time profile of the quark-gluon plasma produced in the Beam Energy Scan:
\begin{itemize}
	\item The temperature and baryon chemical potential vary more slowly around $\eta_s=0$. At higher collision energies, the $\eta_s=0$ region forms a well-known plateau extending two to three units of spatial rapidity. This plateau shrinks with decreasing $\sqrt{s_{NN}}$ and essentially vanishes at $\sqrt{s_{NN}}=19.6$~GeV.
	\item Beyond the central ($\eta_s=0$) plateau, the average temperature decreases rapidly with spatial rapidity while the ratio $\mu_B/T$ increases rapidly. 
	\item The baryon chemical potential at $\eta_s\approx 0$ in $\sqrt{s_{NN}}=19.6$~GeV collisions corresponds approximately to the baryon chemical potential at $\eta_s\approx \pm 2.5$ in $\sqrt{s_{NN}}=200$~GeV collisions.
\end{itemize}

Note that the average temperature in the central plateau ($\eta_s\sim 0$) is not significantly higher at  $\sqrt{s_{NN}}=200$~GeV than it is at $\sqrt{s_{NN}}=19.6$~GeV. The reason for this is in part physical and in part a consequence of the definition of average used in this work.  Because the quark-gluon plasma expands rapidly, its low temperature spacetime volume is much larger than its higher temperature volume. This means that averages of temperature tend to be dominated by the lower temperature regions of the plasma no matter what the center-of-mass energy $\sqrt{s_{NN}}$ is. This is a physical effect, and it is known that a large number of photons are produced at lower temperature (late time) by this large spacetime volume. On the other hand, the cut-off used in the average, $\epsilon(X)>0.16\textrm{ GeV/fm}^3$, excludes very low energy density regions from the average, and thus limits how low the average temperature can be. As long as the cut-off is used consistently across all systems, this is not an issue and different $\sqrt{s_{NN}}$ can still be compared consistently\footnote{Temperature averages always require a weight or a cut-off to suppress the contribution of the low temperature regions of the plasma, which essentially extend to infinity. Different prescriptions are used in the literature and comparisons between different results must be made with care.}.

\begin{figure}
	\centering
	$\vcenter{\hbox{\includegraphics[width=0.35\linewidth]{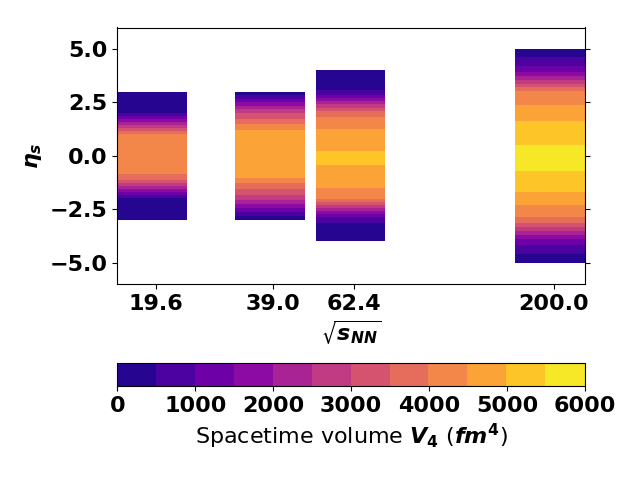}}}$
	\hspace*{.01in}
	$\vcenter{\hbox{\includegraphics[width=0.3\linewidth]{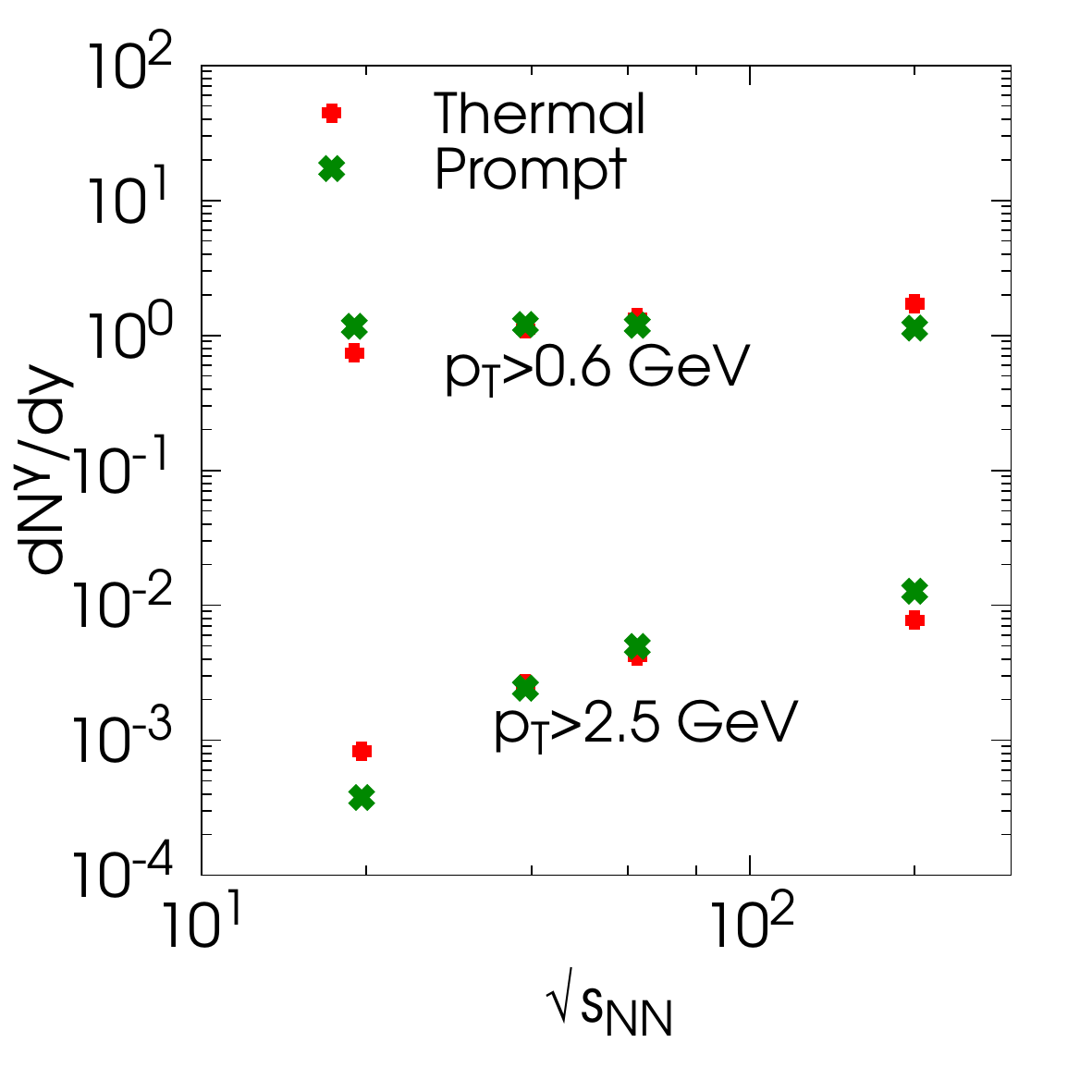}}}$
	\hspace*{.01in}
	$\vcenter{\hbox{\includegraphics[width=0.31\linewidth]{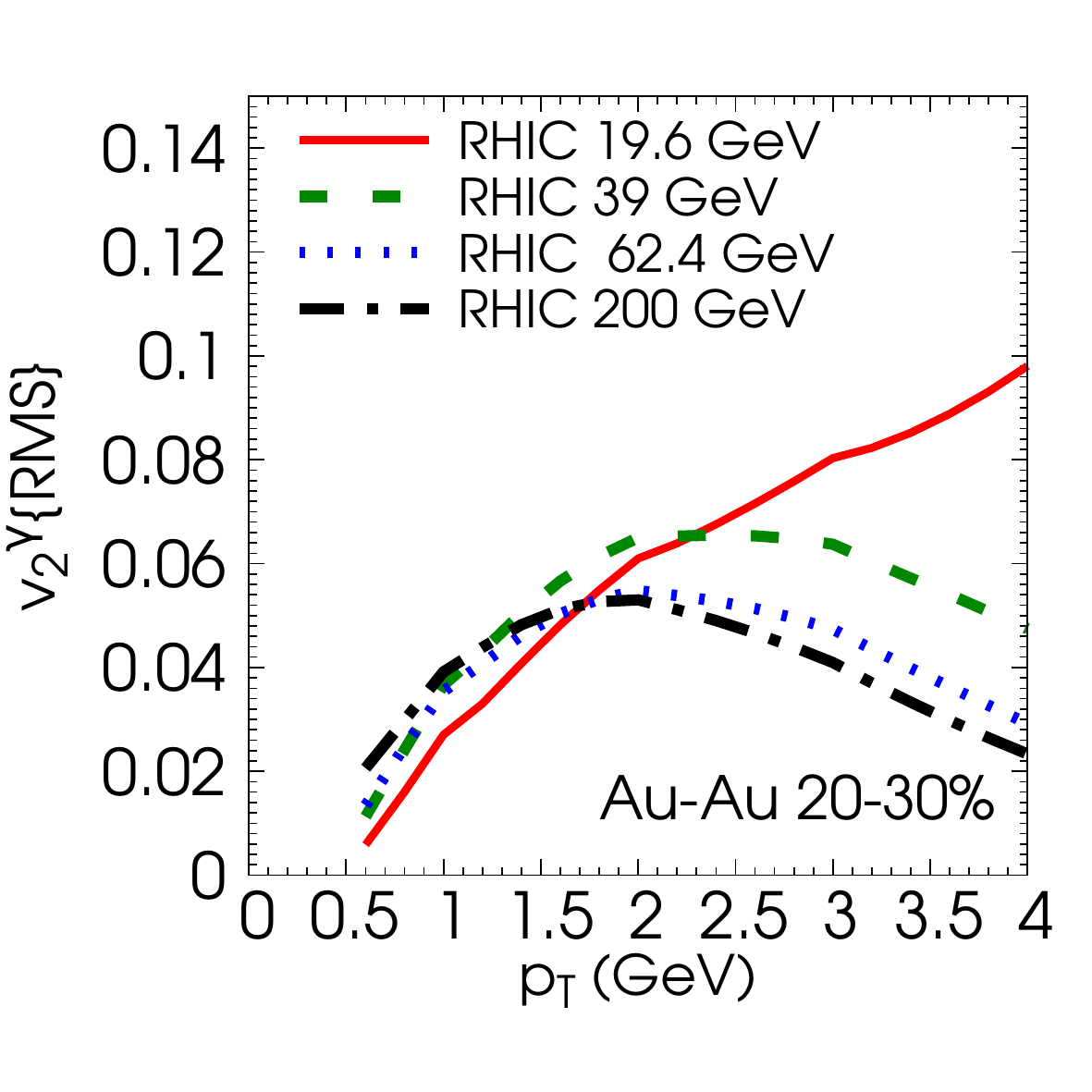}}}$
	\caption{Left: Spacetime volume of plasma with $\epsilon(X)>0.16\textrm{ \footnotesize GeV/fm}^3$ as a function of center-of-mass energy $\sqrt{s_{NN}}$ and spatial rapidity $\eta_s$. Middle: photon multiplicity at different $\sqrt{s_{NN}}$ with two $p_T^\gamma$ cuts. Right: Direct photon (thermal + prompt) $p_T^\gamma$-differential $v_2$ coefficient in 20-30\% Au+Au collisions at four collision energies.}
	\label{fig:rapidity_cm_V4}
\end{figure}

In Figure~\ref{fig:rapidity_cm_V4} (left panel) we show the spacetime volume $V_4$ of the plasma as defined with the energy density cut-off $\epsilon(X)>0.16\textrm{ GeV/fm}^3$. Photon emission is only significant in regions where both $V_4$ and the temperature/$\mu_B$ are large. The rapid decrease of both the spacetime volume and the temperature of radiating plasma as a function of spatial rapidity $\eta_s$ implies that most photons are produced at $\eta_s\sim 0$, as can be reasonably expected. 
%Isolating photons produced in other regions of the plasma require kinematic cuts on the photon momentum.
%, which we discuss in the following section.

\section{Photon production in $\sqrt{s_{NN}}$ and momentum rapidity}

Thermal photon production is calculated by convoluting the spacetime profile of the medium from hydrodynamics with a photon emission rate calculated from first principles:
$
k \frac{d^3 N}{d^3 k}=\int d^4 X k \frac{d\Gamma_\gamma}{d^3 k}\left(T(X),\mu_B(X), \ldots \right)
$

The photon emission rate per spacetime volume $k d\Gamma_\gamma/d^3 k$ has been calculated at very high temperature ($\alpha_s(T)\ll 1$) and finite $\mu_B/T$ using perturbative QCD in Refs.~\cite{Traxler:1994hy,Gervais:2012wd}. The photon rate from an interacting fluid of hadrons at temperatures $T\sim 100$~MeV and $\mu_B/T\sim 0-3$ is also known~\cite{Heffernan:2014mla,Turbide:2003si}. In this work the photon emission rate of the strongly-coupled quark-gluon plasma is calculated by using this hadronic rate for temperatures below 180~MeV while the perturbative rate is used above this temperature~\cite{Paquet:2015lta}. We expect this prescription to provide a photon emission rate for the strongly-coupled quark-gluon plasma that has a sound temperature and $\mu_B$ dependence. We nevertheless emphasize that this prescription should be continuously revisited, both at zero and finite $\mu_B$, as new investigations of the photon emission rates are performed (see Ref.~\cite{Paquet:2016pnt} for references to recent works).

Prompt photons are calculated with next-to-leading order perturbative QCD~\cite{Aurenche:2006vj} as described in Ref.~\cite{Paquet:2015lta}. We note that we use an extrapolation procedure described in Ref.~\cite{Paquet:2015lta} to obtain prompt photons at low $p_T^\gamma$. This extrapolation inevitably introduces a significant uncertainty, especially for low $\sqrt{s_{NN}}$. Given the absence of low $p_T^\gamma$ p+p photon measurements for most collision energies, we proceed with this extrapolation.

The thermal and prompt photon multiplicity are shown as a function of $\sqrt{s_{NN}}$ on the middle panel of Figure~\ref{fig:rapidity_cm_V4}. The multiplicity of higher $p_T^\gamma$ thermal photons ($p_T^\gamma>2.5$~GeV in this example), which are dominantly produced in higher temperature region of the plasma, increases rapidly with $\sqrt{s_{NN}}$. This can be expected from the increasing volume of high temperature plasma found at higher collision energy. The multiplicity of lower $p_T^\gamma$ thermal photons ($p_T^\gamma>0.6$~GeV) increases much more slowly, indicating that there is already a sufficiently large volume of lower temperature plasma to produce thermal photons even in low $\sqrt{s_{NN}}$ collisions.

Interestingly the prompt photon multiplicity at higher $p_T^\gamma$ increases more rapidly with $\sqrt{s_{NN}}$ than the thermal photon multiplicity. We understand this to be a consequence of suppression of prompt photon production at large values of $x_T^\gamma=2 p_T^\gamma/\sqrt{s_{NN}}$. With $p_T^\gamma\sim 2$~GeV and $\sqrt{s_{NN}}=19.6$~GeV, $x_T^\gamma\approx 0.2$. It is known that the production of prompt photon decreases more rapidly with $x_T^\gamma$ when such large values of $x_T^\gamma$ are probed (see e.g. Ref.~\cite{Adare:2012yt}).
%This suppression can be understood as follow. Prompt photons scale approximately with as 
%\begin{equation}
%k \frac{d^3 N}{d^3 k}  \sim \frac{F\left(2 p_T^\gamma/\sqrt{s_{NN}}\right)}{\sqrt{s_{NN}}^{a}}
%\end{equation}
%where $F(x_T=2 p_T^\gamma/\sqrt{s_{NN}})$ is a monotonically decreasing function of the transverse momentum fraction $x^\gamma_T$ and $a$ is an exponent of order $4-5$ when $\sqrt{s_{NN}}$ is in GeV.
For prompt photon production at \emph{lower} $p_T^\gamma$, our prompt photon multiplicity remains relatively constant with $\sqrt{s_{NN}}$. We believe this is evidence that our calculation of low $p_T^\gamma$ photons, extrapolated as described in Ref.~\cite{Paquet:2015lta}, is being pushed beyond its regime of validity. Low $p_T^\gamma$ photon measurements in proton-proton collisions will be essential to address this challenge. Nevertheless we believe the rapid change of higher $p_T^\gamma$ prompt photons with $\sqrt{s_{NN}}$ to be physical in origin, and that it can provide a window into thermal photons in low $\sqrt{s_{NN}}$ collisions.

%\begin{figure}
%	\centering
%	\includegraphics[width=0.33\linewidth]{v2_thermal_prompt_photons_AuAu_BES_cent2030_196_39}
%	\hspace*{.4in}
%	\includegraphics[width=0.33\linewidth]{v2_thermal_prompt_photons_AuAu_BES_cent2030_624_200}
%	\caption{$v_2\{\textrm{RMS}\}$ of direct photons (thermal+prompt) at four different $\sqrt{s_{NN}}$ }
%	\label{fig:v2}
%\end{figure}

Finally we show the momentum anisotropy $v_2$ of photons in the right panel of Fig.~\ref{fig:rapidity_cm_V4}. The photon $v_2$ is large at all $\sqrt{s_{NN}}$. The observations of a large $v_2$ at low $\sqrt{s_{NN}}$ was also made earlier in hadron measurements~\cite{Adamczyk:2013gw,Snellings:2014vqa}. With photons, there is however an additional effect not encountered with hadrons: while the \emph{thermal} photon $v_2$ generally correlates well with the hadronic $v_n$, the \emph{direct} photon $v_2$ also incorporates the effect of prompt photons, which suppress the thermal photon $v_2$. Since less prompt photons are produced at higher $p_T^\gamma$ and low $\sqrt{s_{NN}}$ (i.e. large $x_T^\gamma$ discussed above), the $v_2$ of direct (thermal+prompt) photon $v_2$ can reach large values, as observed at high $p_T^\gamma$ for $\sqrt{s_{NN}}=19.6$~GeV. While further calculations will be necessary to investigate the effect of viscosity on such high $p_T$ thermal photons, we believe the interesting aspects of photon production at low $\sqrt{s_{NN}}$ supports investing more resources into theoretical and experimental investigations of photon production at low collision energies.
%This is not unexpected given that the $v_2$ of hadrons is also large for a wide range of $\sqrt{s_{NN}}$~\cite{Snellings:2014vqa}. The direct photon $v_2$ is also be helped at low $\sqrt{s_{NN}}$ by the suppression of prompt photons discussed above: the small $v_2$ of prompt photons tends to dilute the $v_2$ of thermal photons. Less prompt photons leads to a larger \emph{direct} (thermal+prompt) photon $v_2$.



%% References

\textbf{Acknowledgements}
This work was supported by the U.S. Department of Energy, Office of 
Science, Office of Nuclear Physics  under Award Numbers DE-FG02-05ER41367 (JFP) and DE-SC0012704 (CS) and by the Natural Sciences and Engineering Research Council of Canada.  CG acknowledges support from the Canada Council for the Arts, through its Killam Research Fellowship Program. SM acknowledges funding from The Fonds de recherche du Qu\'ebec - Nature et technologies (FRQ-NT) through the Programme de Bourses d'Excellence pour \'Etudiants \'Etrangers.
Computations were made in part on the supercomputer Guillimin, managed by Calcul Qu\'ebec and Compute Canada and funded by the Canada Foundation for Innovation (CFI), Minist\`ere de l'\'Economie, des Sciences et de l'Innovation du Qu\'ebec (MESI) and FRQ-NT. This research used resources of the National Energy Research Scientific Computing Center, which is supported by the DOE Office of Science under Contract No. DE-AC02-05CH11231.

\bibliographystyle{elsarticle-num}
\bibliography{biblio}

\begin{thebibliography}{10}
\expandafter\ifx\csname url\endcsname\relax
  \def\url#1{\texttt{#1}}\fi
\expandafter\ifx\csname urlprefix\endcsname\relax\def\urlprefix{URL }\fi
\expandafter\ifx\csname href\endcsname\relax
  \def\href#1#2{#2} \def\path#1{#1}\fi

\bibitem{Paquet:2016pnt}
J.-F. Paquet, J. Phys. Conf. Ser. 832~(1) (2017) 012035.

\bibitem{Denicol:2018wdp}
G.~S. Denicol, C.~Gale, S.~Jeon, A.~Monnai, B.~Schenke, C.~Shen\href
  {http://arxiv.org/abs/1804.10557} {\path{arXiv:1804.10557}}.

\bibitem{Shen:2017bsr}
C.~Shen, B.~Schenke, Phys. Rev. C97~(2) (2018) 024907.

\bibitem{Shen:2018pty}
C.~Shen, B.~Schenke, in: {27th International Conference on Ultrarelativistic
  Nucleus-Nucleus Collisions (Quark Matter 2018) Venice, Italy, May 14-19,
  2018}, 2018.
\newblock \href {http://arxiv.org/abs/1807.05141} {\path{arXiv:1807.05141}}.

\bibitem{Shen:2017fnn}
C.~Shen, B.~Schenke, PoS CPOD2017 (2018) 006.

\bibitem{Chatterjee:2013naa}
R.~Chatterjee, H.~Holopainen, I.~Helenius, T.~Renk, K.~J. Eskola, Phys. Rev.
  C88 (2013) 034901.

\bibitem{Shen:2013vja}
C.~Shen, U.~W. Heinz, J.-F. Paquet, C.~Gale, Phys. Rev. C89~(4) (2014) 044910.

\bibitem{Paquet:2015lta}
J.-F. Paquet, C.~Shen, G.~S. Denicol, M.~Luzum, B.~Schenke, S.~Jeon, C.~Gale,
  Phys. Rev. C93~(4) (2016) 044906.

\bibitem{Traxler:1994hy}
C.~T. Traxler, H.~Vija, M.~H. Thoma, Phys. Lett. B346 (1995) 329--334.

\bibitem{Gervais:2012wd}
H.~Gervais, S.~Jeon, Phys. Rev. C86 (2012) 034904.

\bibitem{Heffernan:2014mla}
M.~Heffernan, P.~Hohler, R.~Rapp, Phys. Rev. C91~(2) (2015) 027902.

\bibitem{Turbide:2003si}
S.~Turbide, R.~Rapp, C.~Gale, Phys. Rev. C69 (2004) 014903.

\bibitem{Aurenche:2006vj}
P.~Aurenche, M.~Fontannaz, J.-P. Guillet, E.~Pilon, M.~Werlen, Phys. Rev. D73
  (2006) 094007.

\bibitem{Adare:2012yt}
A.~Adare, et~al., Phys. Rev. D86 (2012) 072008.

\bibitem{Adamczyk:2013gw}
L.~Adamczyk, et~al., Phys. Rev. C88 (2013) 014902.

\bibitem{Snellings:2014vqa}
R.~Snellings, EPJ Web Conf. 97 (2015) 00025.

\end{thebibliography}






% \end{thebibliography}

\end{document}